\documentclass[11pt]{article}

\usepackage{amsmath,amssymb,amsfonts,amsthm}
\usepackage{graphicx} 
\usepackage{authblk}
\usepackage{subcaption} 

\makeatletter
\newcommand*{\transpose}{%
	{\mathpalette\@transpose{}}%
}
\newcommand*{\@transpose}[2]{%
	\raisebox{\depth}{$\m@th#1\intercal$}%
}

\begin{document}

\title{An elastodynamic Willis meta-slab with strong directional-dependent absorption.}

\author[$\dagger$]{Phillip A.\ Cotterill}
\author[$\star$]{David Nigro}
\author[$\dagger$]{William J.\ Parnell}

\affil[$\dagger$]{Department of Mathematics, University of Manchester, Oxford Road, Manchester, M13 9PL, UK}
\affil[$\star$]{Thales UK, 350 Longwater Avenue, Reading, Berkshire, RG2 6GF, UK}

\maketitle

\begin{abstract}
Electromagnetic bi-anisotropy finds an analogy in acoustic metamaterial science as Willis coupling. Its impact and emergence in the field of elastodynamic metamaterials is not as well understood however, given the coupling between compressional and shear waves. Here we discuss the emergence of Willis coupling in heterogeneous elastic slabs embedded in an acoustic fluid. The microstructure of the slab comprises circular cylindrical voids and asymmetry is present via two neighbouring line arrays, each with a repeating void of differing radius. The slab matrix is soft, with Poisson ratio close to $1/2$ so that the voids act as Giant Monopole Resonators, and induce a strong dynamic response at low frequency. The incorporation of Willis constitutive coupling ensures that a unique set of effective material properties can be assigned to the slab up to a maximum frequency defined by the periodic spacing of the voids and the elastic properties of the substrate. Including loss in the elastic medium via its shear modulus induces strong directional-dependent absorption at low frequency, whilst of course maintaining reciprocity.
%
%
%
%
%
%
%
%
\end{abstract}




\section{Introduction}

Metamaterials employ subwavelength microstructures in order to induce novel material properties and thus manipulate waves. Metamaterials are now prevalent in a wide range of fields including acoustics \cite{ma2016acoustic, cummer2016controlling, haberman2016acoustic}, electromagnetism \cite{cui2010metamaterials, zheludev2012metamaterials}, thermodynamics \cite{schittny2013experiments} and mechanics \cite{bertoldi2017flexible}. Advances have also been made in the field of elastodynamics \cite{kadic2013metamaterials, srivastava2015elastic}, where the scenario is often made more complex by mode conversion between compressional and shear waves at boundaries. 

Significantly, the governing equations of the classical fields noted above are all invariant under coordinate transformations, with the exception of elastodynamics \cite{willis2022some, willis1981variational, norris2011elastic}. Indeed, for the equations of elastodynamics to be invariant with respect to coordinate transformations, it transpires that additional effects are required in the governing equations. In metamaterial science, the effect now commonly incorporated to ensure invariance is \textit{Willis coupling}. This is a dynamic constitutive effect coupling strain and momentum fields, as well as stress and velocity fields \cite{willis1981variational, norris2011elastic, Milton06, muhlestein2016reciprocity}. Local Willis constitutive relations can be written in the form
\begin{align}
\boldsymbol{\sigma} &= \mathbf{C}:\mathbf{e}+\boldsymbol{\psi}\cdot\ddot{\mathbf{u}} \label{W1.1}\\
\boldsymbol{\mathcal{M}} &= \boldsymbol{\psi}^{\dagger}\cdot\dot{\mathbf{e}} + \boldsymbol{\rho}\cdot\dot{\mathbf{u}}
\end{align}
where $\boldsymbol{\sigma}$ is the Cauchy stress tensor, $\boldsymbol{\mathcal{M}}$ the momentum density, $\mathbf{e}=\frac{1}{2}(\nabla\mathbf{u}+(\nabla\mathbf{u})^T)$ the strain tensor, $\mathbf{u}$ the displacement vector. $\mathbf{C}$ is the (fourth order) elastic modulus tensor, $\boldsymbol{\rho}$ is the (second order) mass density tensor, and finally $\boldsymbol{\psi}$ is the (third order) Willis coupling tensor. Dagger here indicates the complex conjugate.
Willis constitutive coupling disappears in the static limit and is typically relatively weak in the low-frequency regime for standard (non-resonant) material inhomogeneity. Homogenisation schemes to determine the form of the effective properties resulting from specific microstructures have been developed by numerous authors, e.g.\  \cite{nassar2015willis, torrent2015resonant} and the theory has also been extended to include other multi-physics effects \cite{pernas2020fundamental}.

Willis coupling has been identified as analogous to bi-isotropy in acoustics and bi-anisotropy in electromagnetism. Extensive work has been carried out in the passive acoustics regime, see e.g.\ \cite{groby2021analytical, peng2022fundamentals}. In order to induce strong low-frequency effects it is necessary to employ resonant microstructure. In Muhlestein et al \cite{muhlestein2017experimental} acoustic Willis coupling was measured experimentally for the first time in one dimension, in the configuration of two distinct resonators. Measurements of the reflection from, and transmission through, this configuration were made. It was then shown that a unique set of effective acoustic properties could only be assigned to the medium if Willis constitutive coupling was incorporated. Without this effect, effective acoustic properties were non-unique.

Work on Willis coupling in elastic systems is increasing in number but is still scarce and often limited to equivalent spring-mass models  \cite{qu2022mass}. Liu et al.\ \cite{liu2019willis} incorporated the Willis effect for flexural waves on beams by exploiting a cantilever bending resonance and described an asymmetry to the reflection coefficient. Including dissipation resulted in asymmetric loss. In such conservative systems asymmetry and strong resonance is the key to the presence of the coupling.  In \cite{hao2022experimental} torsional waves were considered experimentally, with coupling induced again via asymmetry of resonators. Scattering in an asymmetric elastic waveguide was also recently considered in \cite{mcintosh2024asymmetric} using approximations to the local resonances of the inclusions comprising the waveguide system. 

With a view on the more general elastodynamic system and access to the larger number of free parameters that could be tuned in the full elastodynamic regime, here we return to an analogous system considered by Muhlestein et al. \cite{muhlestein2017experimental} but now in the setting of elastodynamics. We introduce the configuration of a \textit{meta-slab} of the sort considered in Ivansson \cite{ivansson2006sound} where a strong dynamic response is ensured by embedded voids in a soft medium so that the Giant Monopole Resonance (GMR) is present \cite{meyer1958pulsation, cotterill2022deeply}. Willis coupling arises due to the fact that we distribute this microstructure asymmetrically inside the slab domain. 

For the GMR to occur we require the meta-slab matrix to have bulk modulus $\kappa$ and shear modulus $\mu$ satisfy $\mu\ll\kappa$. In this case a strong dynamic effect results at low frequency \cite{meyer1958pulsation, ivansson2006sound}. The case of the effect of interaction on the GMR by closely-space voids was discussed recently in \cite{cotterill2022deeply}, where it was shown that interaction affects the resonant frequency significantly but does not modify the amplitude. This is in contrast to the interaction of dipole resonators of the classical form \cite{liu2000locally} which do not modify their resonant frequency even when in close proximity, although their resonant amplitude is enhanced significantly in contrast to the void case \cite{touboul2022enhanced}.

In order to study the emergence of Willis coupling in this elastodynamic system, we embed the meta-slab inside an acoustic medium (supporting only compressional waves). Incident compressional waves are considered on both sides of the slab and from the reflection and transmission coefficients we determine the associated effective properties, including the necessary component of Willis coupling via an inverse procedure. We solve the problem inside the slab via a hybrid analytical-numerical procedure, which we detail in the appendix, and equate this to an effective, homogeneous medium. This results in our determination of the effective properties of the meta-slab, and in particular, when we include loss we show that asymmetry leads to the effect of directional-dependent absorption. This configuration, and microstructual resonance, illustrates the critical role that shear has in the elastodynamic scenario in a distinct manner from acoustics.

In Section \ref{Sec2} we provide an overview of the problem including the configuration and discussion of the formulation of the Willis medium scattering problem. The homogenisation methodology, regarding how we link the solution of the actual problem with microstructure, to the effective Willis medium problem is then given in Section \ref{Sec3}. Results follow in Section \ref{Sec4}, detailing a specific case of asymmetry both without and with loss in the meta-slab matrix. We provide conclusions in Section \ref{Sec5}.

\section{Problem Overview} \label{Sec2}

\begin{figure}
\begin{subfigure}[h]{0.4\linewidth}
\includegraphics[width=\linewidth]{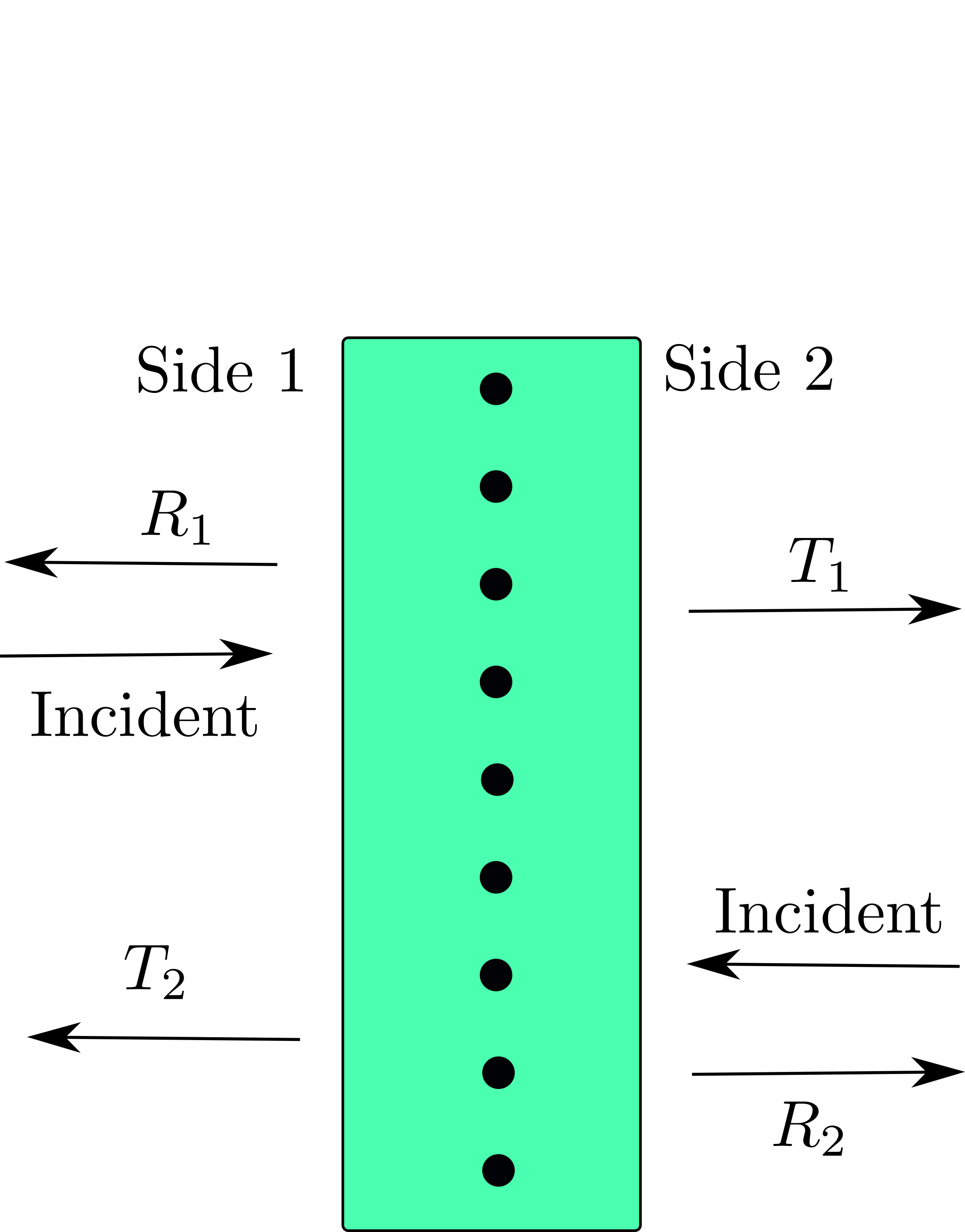}
\caption{}
\end{subfigure}
\hfill
\begin{subfigure}[h]{0.4\linewidth}
\includegraphics[width=\linewidth]{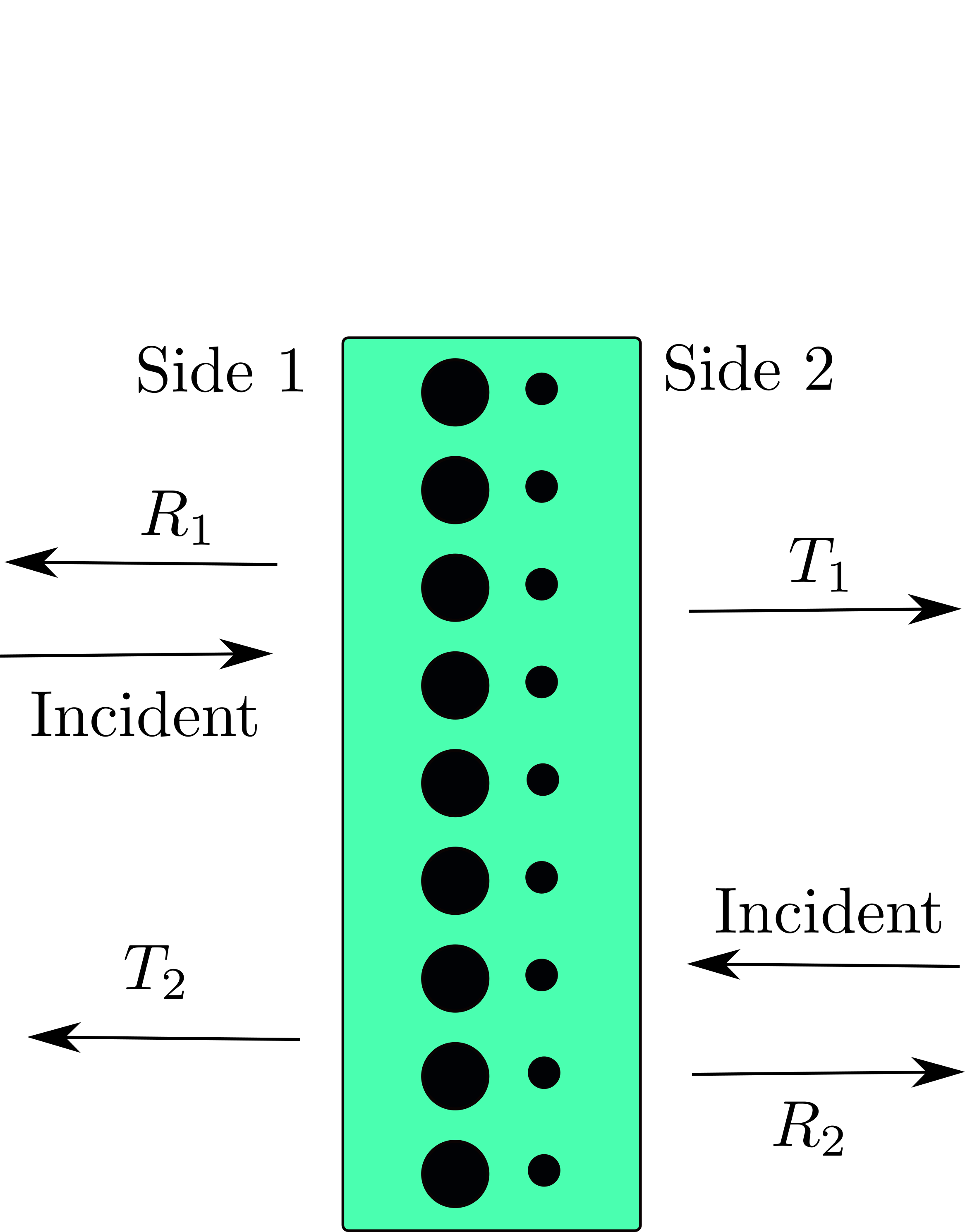}
\caption{}
\end{subfigure}%
\caption{Elastic slab with (a) symmetric and (b) asymmetric microstructure consisting of circular cylindrical voids distributed on line arrays inside the slab. In the asymmetric case the configuration consists of two arrays of voids with different radii. Incident fields from the left and right give rise to reflection/transmission coefficients with subscripts 1 and 2 respectively.}
\label{config}
\end{figure}

With reference to Figure \ref{config}, we study a two-dimensional configuration involving an elastic slab occupying the domain $0\leq x\leq h$, of infinite extent in the $y$ and $z$ directions, and having internal inhomogeneous structure. This slab resides within an acoustic medium occupying $x< 0$ and $x>h$, with known properties  here defined by the compression modulus $M_f$ and density $\rho_f$. Waves are assumed time-harmonic with the associated fields of the form $\overline{f}(\mathbf{x},t)= f(\mathbf{x})e^{-i\omega t}$, where $\mathbf{x}$ is the spatial coordinate, $t$ is time and $\omega$ is angular frequency. We consider the configurations as illustrated in Figure \ref{config} but critically although the configuration in Figure \ref{config}(a) has been studied extensively, the asymmetric case in Figure \ref{config}(b) has not. Initially we assume that all media are lossless, although later we incorporate loss in the slab to accommodate absorption. For the specific configuration studied we assume that the acoustic medium has impedance $Z_f=Z_m$ and mass density $\rho_f=\rho_m$ where properties with subscript $m$ are the properties of the matrix of the slab. We assume that the inhomogeneous slab responds as an effective homogeneous medium with effective compression modulus $M_*$ and mass density $\rho_*$. As previously noted, in the acoustic case it has been shown that in the most general scenario, an effective medium response for the slab is required to be of the Willis type, and thus we assign the slab with an effective Willis coupling parameter $\psi_*$. As we illustrate in Figure \ref{config} in order to determine the effective properties of the slab we are required to consider an incident field on both sides of the slab. We talk further about this aspect after we have discussed the effective response of the Willis slab.

Given the normally incident compressional field, the problem is quasi-one-dimensional with dependence only on $x_1=x$ and $t$. The displacement field takes the form $\mathbf{u}=(\bar{u}(x,t),0,0)$ and the only non-zero stress component is $\sigma_{11}=\bar{\sigma}(x,t)$. In the instance where insonification is from the left, the pressure field in the acoustic medium is $\overline{p}(\mathbf{x},t)=p(x)e^{-i\omega t}$ where
\begin{align}
p(x) &= \begin{cases}
p_i e^{ik_fx}+p_r e^{-ik_f x}, & x\leq 0, \\
p_t e^{ik_f(x-h)}, & x\geq h
\end{cases}
\end{align}
and we define $R_1=p_r/p_i, T_1=p_t/p_i$. Modified, equivalent notation is employed when the insonification is from the right, with subscripts $2$ as noted in Fig.\ \ref{config}. The associated normal displacements in the fluid are of the form
\begin{align}
u(x) &= \begin{cases}
\dfrac{i}{\rho_fc_f^2k_f}(p_i e^{ik_fx}-p_r e^{-ik_f x}), & x\leq 0, \\
\dfrac{i}{\rho_fc_f^2k_f}p_t e^{ik_f(x-h)}, & x\geq h.
\end{cases}
\end{align}

Given the time-harmonic response and for the purposes of identifying properties of an effective homogenised medium, we assume that the slab responds as an effective Willis medium with dynamic constitutive expressions
\begin{align}
\sigma_* &= M_*\frac{du_*}{dx} - \omega^2\psi_* u_*, \label{W1}\\
\mathcal{M}_* &= -i\omega\psi_*^{\dagger}\frac{du_*}{dx}-i\omega\rho_* u_*. \label{W2}
\end{align}


In the effective elastic slab we assign the properties $\mathbf{C}_*$ (a fourth order elastic modulus tensor) and $\rho_*$ (the density) to the medium. The time-harmonic displacements $\overline{\mathbf{u}}_*(\mathbf{x},t)=\mathbf{u}_*(\mathbf{x})e^{-i\omega t}$ are governed by
\begin{align}
\textnormal{div}\boldsymbol{\sigma}_* + i\omega \mathbf{\mathcal{M}}_* &= 0,
\end{align}
with $\overline{\boldsymbol{\sigma}}_* = \boldsymbol{\sigma}_*e^{-i\omega t}$ and $\overline{\boldsymbol{\mathcal{M}}}_* = \boldsymbol{\mathcal{M}}_*e^{-i\omega t}$. In general $\boldsymbol{\sigma} = \mathbf{C}_*:\mathbf{e}_*$ where $\mathbf{e}_*=\frac{1}{2}(\nabla\mathbf{u}_*+(\nabla\mathbf{u}_*)^T)$. 
We have also used the notation  $(C_*)_{1111}=M_*=\lambda_*+2\mu_*$ and $(\boldsymbol{\psi}_*)_{111}=\psi_*$. Given that the incident field is normal to the interface, the zero shear traction condition is automatically satisfied for the effective slab problem.  From \eqref{W2} we have
\begin{align}
\mathcal{M}_* &= -i\omega \left(\psi_*^{\dagger} \frac{du_*}{dx} +\rho_* u_*\right).
\end{align}

The displacement in the effective slab is thus governed by
\begin{align}
M_* \frac{d^2 u_*}{dx^2} + \omega^2(\psi_*^{\dagger}-\psi_*)\frac{du_*}{dx}+\rho_*\omega^2 u_* &= 0
\end{align}
and given that here we consider passive media, $\psi_*$ is real \cite{muhlestein2016reciprocity} so that this becomes
\begin{align}
M_* \frac{d^2 u_*}{dx^2} +\rho_*\omega^2 u_* &= 0.
\end{align}
This has plane wave solutions
\begin{align}
u_* &= S_R e^{ik_* x} + S_L e^{-ik_* x}
\end{align}
with $k_*^2=\rho_*\omega/M_*$ and associated stress
\begin{align}
\sigma_* &= M_*k_*(S_R e^{ik_* x} - S_L e^{-ik_* x}).
\end{align}
The Willis coupling parameter $\psi_*$ does not arise in the boundary value problem above, although it does affect the field at interfaces. Further, it transpires that in order to determine unique effective properties $M_*$ and $\rho_*$ from the inverse problem outlined below, one requires the presence of this Willis parameter. For amplitudes $\Sigma$ and $V$ associated with stress and velocity in the slab, one can derive an impedance condition
\begin{align}
-\frac{\Sigma}{V} &= \pm \rho_*c_* +i \omega \psi_* \\
 &= Z_*(\pm 1+i W) \label{impconds}
\end{align}
with $Z_*=\rho_*c_*$ and $W=\omega\psi_*/\sqrt{\rho_*M_*} = \omega\psi_*/(\rho_*c_*)$.

We note that continuity conditions for the effective slab problem on $x=0$ and $x=h$ are
\begin{align}
p &= -\sigma_*, & u &= u_*. \label{contconds}
\end{align}
It is precisely these continuity conditions \eqref{contconds} and the effective impedance conditions \eqref{impconds} that allow us to determine the Willis coupling parameter, coupling the effective homogenised response to the exact scattering problem.

\section{Homogenisation methodology}\label{Sec3}
The idea here, in order to identify effective properties of the slab medium is to solve the exact multiple scattering problem associated with some given microstructure. This results in values for the reflection and transmission coefficients, $R$ and $T$ respectively. As described in  \cite{fokin2007method} one can then determine $M_*$ and $\rho_*$ from $R$ and $T$. As already described for asymmetric microstructure this retreival process results in non-unique effective properties. Muhlestein et al.\ \cite{muhlestein2017experimental} extended the method in \cite{fokin2007method} to incorporate Willis material behaviour in the acoustics scenario, in order to deduce the Willis coupling term from modified $R$ and $T$ coefficients. Here we employ an analogous technique, also employing a method developed by Brazier-Smith and Clarke \cite{brazier2008pulse}.

Let us now summarise this parameter retrieval approach, given $R$ and $T$, before explaining how we determine the reflection and transmission coefficients in our elastodynamic regime via a quasi-analytical multiple scattering methodology.



\subsection{Retrieval of material parameters}

Let us consider an elastic slab as illustrated in Figure \ref{config}. To fix ideas, consider the case in figure \ref{config}(b). The slab contains two linear arrays of circular cylindrical voids whose axes are infinitely long and lie parallel to the $z-$axis.  Each linear array contains an infinite number of identical voids that are periodically spaced along the $y-$axis with separation $d$; they are embedded in an otherwise homogeneous elastic substrate of thickness $h$.  The voids are centred at $(x = x_1,y = nd)$ and $(x = x_2, y = nd)$, for $n\in\mathbb{Z}$, with radii $a_1$ and $a_2$ respectively.

As noted above we investigate the response of the slab when a plane pressure wave, propagating in the fluid, is normally incident on each of its \textit{two} surfaces.  With the slab geometry described above this type of excitation reduces the effective slab problem to two dimensions (plane-strain) and, at low frequencies, such a system can be regarded as a one-dimensional Willis material \cite{muhlestein2017experimental}.  

The incident field gives rise to reflection and transmission coefficients at the slab boundaries, and whilst the transmission coefficients are independent of which side is insonified (due to reciprocity), in a Willis material the reflection coefficients are different due to the asymmetry of the slab, which means the slab cannot be represented as a simple homogeneous material.  For this one-dimensional problem, Muhlestein shows that the common transmission coefficient, $T$, is given by 
\begin{equation}\label{eq:trans}
  T = \frac{2z}{2z\cos\alpha - \text{i}\sin\alpha\left[1 + z^2(1+W^2)\right]}.
\end{equation}
whilst the reflection coefficients from the two sides are given by
\begin{equation}\label{eq:reflect}
  R_1 = \frac{\text{i}\sin\alpha\left[1-z^2(1+W^2)-\text{i}2zW\right]}{2z\cos\alpha - \text{i}\sin\alpha\left[1 + z^2(1+W^2)\right]},\quad R_2 = \frac{\text{i}\sin\alpha\left[1-z^2(1+W^2)+\text{i}2zW\right]}{2z\cos\alpha - \text{i}\sin\alpha\left[1 + z^2(1+W^2)\right]}.
\end{equation}
with $R_1$ denoting the reflection coefficient from side 1 due to an incident plane wave propagating from $x = -\infty$, and $R_2$ is the reflection coefficient from side 2 when the plane wave propagates from $x = \infty$.

In equations \eqref{eq:trans} and \eqref{eq:reflect}: $z = \rho_*c_*/(\rho_fc_f)$, where $\rho_*$ and $c_*$ are the effective density and compression wave-speed of the slab and $\rho_f,\,c_f$ denote the same parameters in the fluid; $\alpha = \omega h/c_*$ is the phase shift of the effective elastic, compressional wave across the slab; and $W = \omega\psi_*/(\rho_*c_*)$ is the asymmetry parameter with $\psi_*$ being the one-dimensional Willis parameter. Concerning $W$ and $\psi_*$, if the slab is replaced by its mirror image about $x = h/2$ (see Figure \ref{config}) then $R_1 \rightarrow R_2$ and vice versa.  From \eqref{eq:reflect}, we see that this is equivalent to reversing the sign of $W$ and hence $\psi_*$.  Thus, these parameters depend upon the orientation of the slab within the chosen coordinate system.

Given the scattering coefficients $R_1,\,R_2$ and $T$, it is possible to invert the formulae above to obtain the effective material properties of the slab.  For non-Willis materials, the inverse method is described in \cite{fokin2007method}, a method that was adapted to incorporate Willis coupling in \cite{muhlestein2017experimental}. Thus, we find
\begin{equation}\label{eq:phaseandimpedance}
    \alpha = -\text{i}\log\left(\mathcal{C} \pm\sqrt{\mathcal{C}^2 - 1}\right) +2n\pi,\quad z = \pm2T\frac{\sqrt{\mathcal{C}^2 - 1}}{(1-R_1)(1-R_2)-T^2},
\end{equation}
where
\begin{equation}\label{eq:Trig}
  \mathcal{C} = \cos\alpha = \frac{1 + T^2 - R_1 R_2}{2T},
\end{equation}
and $n$ is an integer.  In addition we calculate the asymmetry parameter, $W$, via:
\begin{equation}\label{eq:asym}
  zW =\frac{\text{i}(R_2 - R_1)}{(1-R_1)(1-R_2) - T^2}.
\end{equation}
In \eqref{eq:phaseandimpedance}, the term $2n\pi$ of $\alpha$ reflects the ambiguity of the inverse trigonometric functions and arises from the branch cut of the $\log$ function, which is usually taken along the negative real axis.  We return to the determination of $n$ later, but at low frequencies it is safe to assume that it is zero, when we see that the sign ambiguity of the square root function leads to $\alpha$ and $z$ switching signs, noting that the product $zW$ is unambiguous.  Provided the relative signs of $\alpha$ and $z$ are correct, the scattering coefficients are insensitive to their actual signs but if the relative signs are wrong, \eqref{eq:trans} and \eqref{eq:reflect} will return incorrect values.  This provides a useful consistency check, but to determine the actual signs of $\alpha$ and $z$ another criterion is required. Fokin et al. \cite{fokin2007method} for example, require that for a \textit{passive} meta-material, $\Re(z) > 0$, which we find works well for some circumstances but not all.  For example, as frequency varies in an undamped system, the periodic configurations described above exhibit a pass/stop band structure.  Within pass bands $\alpha$ and $z$ are real, and the method developed in \cite{fokin2007method} suffices, but in stop bands $z$ is purely imaginary, and so this method cannot be used.  Further, in a stop band $\alpha$ becomes complex and has the form $\alpha = m\pi + \text{i}\delta$, where $m$ is an integer and $\delta$ is real.  Thus $\mathcal{C}$ remains real but has a magnitude greater than 1.  However, once $z$ is determined, the correct sign of $\alpha$ follows (from the consistency check), with $n$ determined by the requirement that $\alpha$ should be a continuous function of increasing frequency; $W$ is obtained from \eqref{eq:asym}.

Brazier-Smith and Clark \cite{brazier2008pulse} write $\alpha$ as
\begin{equation}\label{eq:phase}
    \alpha = -\text{i}\log\left(\mathcal{C} -\sqrt{\mathcal{C} + 1}\sqrt{\mathcal{C} - 1}\right) +2n\pi.
\end{equation}
Given the usual convention that the branch cut of the complex square root function is taken along the negative real axis, the branch cut in \eqref{eq:phase} lies along the real $\mathcal{C}$ axis between the two branch points at $\mathcal{C} = \pm 1$. From the consistency check described above, the corresponding value for the impedance is found to be
\begin{equation}\label{eq:impedance}
    z = 2T\frac{\sqrt{\mathcal{C} + 1}\sqrt{\mathcal{C} - 1}}{(1-R_1)(1-R_2)-T^2}.
\end{equation}
This method works well if the system is damped because then $\mathcal{C} = \cos\alpha$ acquires an imaginary part, which moves it away from the branch cut and the choice of cut ensures that $\Im(\alpha) > 0$.\footnote{$\mathcal{C} \pm \sqrt{\mathcal{C} + 1}\sqrt{\mathcal{C} - 1}$ are inverses of each other, and with damping, one of the signs will have a magnitude less than 1.}  This method also works within stop bands of an undamped system where $\mathcal{C}$ is real, but has a magnitude greater than one. However, in an undamped pass band $\mathcal{C}$ lies on the branch cut, making it sensitive to small errors in the scattering coefficients, which in the cases studied here lead to multiple crossings of the cut.

To determine the value of $n$ used in \eqref{eq:phase}, Brazier-Smith and Clark \cite{brazier2008pulse} use a phase unwrapping technique, assuming that at sufficiently low frequencies $n = 0$. Then, as the frequency increases, the locus of $e^{\text{i}\alpha}$ is followed in the complex plane.  Every time $e^{\text{i}\alpha}$ crosses the negative real axis in a counterclockwise direction $n$ is increased by 1 and every time the crossing is made in a clockwise direction, $n$ is reduced by 1. The inverse methods developed by either \cite{muhlestein2017experimental}, or  \cite{brazier2008pulse} can be used when damping is present.

Here, we begin with the Brazier-Smith and Clark method of \eqref{eq:phase} and \eqref{eq:impedance} (with $n=0$) but if the system is undamped we check the sign of $\Re(z)$ in the pass bands and, if necessary, reverse the signs of both $z$ and $\alpha$.  The correct value of $n$ is then determined by the phase unwrapping technique.

Having determined $\alpha,\,z$ and $W$, the effective elastic parameters of the slab are thus given by
\begin{equation}
    c_* = \frac{\omega h}{\alpha},\quad Z_* = \rho_*c_* = \rho_fc_fz,\quad M_* = Z_* c_*,\quad \rho_* = \frac{Z_*}{c_*},\quad \psi_* = \frac{Z_*W}{\omega}.
\end{equation}

Before leaving this section, we note that the parameter $zW = \omega\psi_*/(\rho_fc_f)$ is essentially a non-dimensional Willis parameter that should be real for a passive meta-material embedded in an undamped fluid.  This is a useful sanity check that is used in the following results section.  Finally, $M_*$ and $\rho_*$ are not affected by the ambiguity in the signs of $\alpha$ and $Z_*$ provided that their relative signs are correct.

\subsection{Approach to finding reflection and transmission coefficients}
In \cite{muhlestein2017experimental}, the scattering coefficients were obtained by experimental measurement.  Here we use a hybrid analytic/numerical method in which multiple scattering theory is used to define system matrices that connect the scattering coefficients within the slab.  The latter arise from interactions between the fields scattered by the circular inclusions and their reflections at the slab boundaries, with the boundary conditions of continuity of stress and normal displacement being applied at the latter.  It would also be possible to solve this problem using the finite element method, or any other numerical scheme as one prefers.  Further details of the method used here are given in the Appendix.

Regardless of the method employed to determine the scattering coefficients of the slab, the periodic nature of the scattering problem means that the fields (pressure, compression and shear) can be represented by an infinite series of plane waves (Bloch waves), some of which will be propagating (cut-on) whilst the rest decay.  The number of cut-on modes increases with increasing frequency; but at normal incidence there will always be at least one cut-on mode of each wave type.  The homogenization schemes described herein are restricted to frequencies below the cut-on values of higher-order modes within either the slab or the fluid, that is, there must be only a single propagating mode of each wave type.

We now consider the implementation of the methods above in order to deduce the wave-carrying properties of meta-slabs with some specific asymmetric microstructure. We deduce their associated effective properties and reflection and transmission coefficients, as well as absorption characteristics when loss is introduced into the matrix.

\section{Results} \label{Sec4}

The case of reflection and transmission from elastic slabs that possess symmetric microstructure has been studied extensively, and often with a methodology that determines the phononic band structure of the structure in question, see e.g.\  \cite{psarobas2000scattering, platts2002two}. The GMR case associated with a soft matrix is known to give rise to a strong low frequency response below the classical resonances that arise due to periodicity and that lead to the classical band gap effect in periodic media \cite{ivansson2006sound, sharma2019sound}. 

Given the extensive discussion of the symmetric microstructure scenario and the fact that it does not give rise to Willis coupling, here we instead focus on asymmetric microstructure. The system we consider is that illustrated in Fig.\ \ref{config}(b), i.e.\ two embedded periodic linear arrays of voids.  The acoustic medium is impedance and density matched to the slab matrix, that is $Z_f/Z_m = 1, \rho_f/\rho_m = 1$, noting again that a subscript $m$ refers to the matrix medium property. We also take a rigidity parameter $\Re(\mu_m)/M_m=0.01$. We will introduce the measures of loss below when we come to discuss the results associated with this scenario.  

With regard to the geometry, the left line array has voids of radius $a_1=a$ with spacing $d_1=d=2.5a$ and the right line array has voids of radius $a_2=a/2$ and spacing $d_2=2.5a$. We choose a slab thickness $h=10a$ which, with rigidity set to $0.01$, then also gives $k_p h=k_sa$. The separation between the arrays is $3a$. The left array has its centre-line at $x=4.4a$ and the centre-line of the right array is at $x=7.4a$, which gives a combined `centre of mass' in the middle of the slab.

In the following we always operate below any higher-order cut on modes so that there is only one propagating mode for compression and shear, but given that we consider normal incidence there is no propagating shear wave in the effective slab. This means that we can uniquely define the effective properties of the slab. As should be the case in this regime these effective properties are also independent of the fluid chosen outside the meta-slab.

\subsection{Meta-slab without loss}

\begin{figure}[h!]
\begin{subfigure}[h]{0.5\linewidth}
\includegraphics[width=\linewidth]{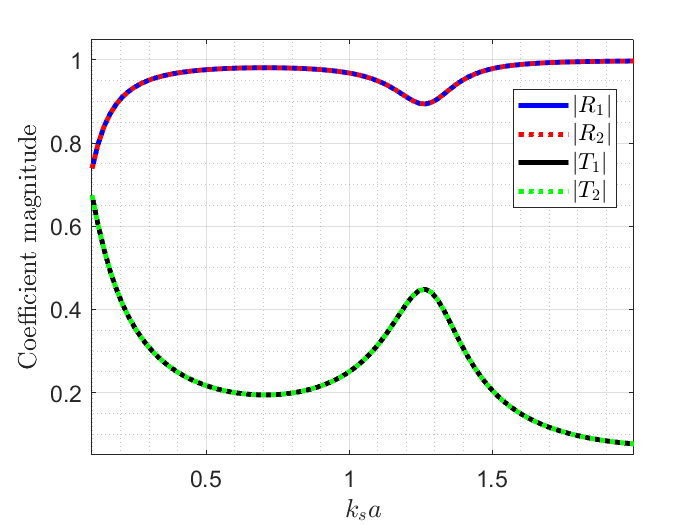}
\caption{}
\end{subfigure}
\hfill
\begin{subfigure}[h]{0.5\linewidth}
\includegraphics[width=\linewidth]{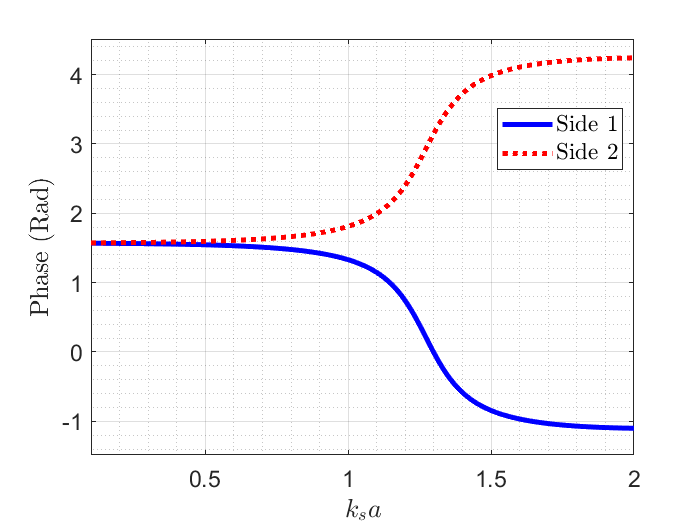}
\caption{}
\end{subfigure}
\caption{Plot of (a) $|R_1|, |R_2|, |T_1|$ and $|T_2|$ as functions of $k_s a (=k_ph)$, and (b) the phase of $R_1$ and $R_2$ relative to $T_1\backslash T_2$, again as functions of $k_s a$ when there is no loss in the matrix medium of the slab.
}
\label{nolossRTphase}
\end{figure}

We first consider the case without loss.  In Fig.\ \ref{nolossRTphase} we plot the magnitude of the reflection and transmission coefficients and the associated phase of the reflection coefficients, noting that due to reciprocity $T_1=T_2$. We see an enhanced transmission.

\begin{figure}[htbp]
\begin{subfigure}[h]{0.5\linewidth}
\includegraphics[width=\linewidth]{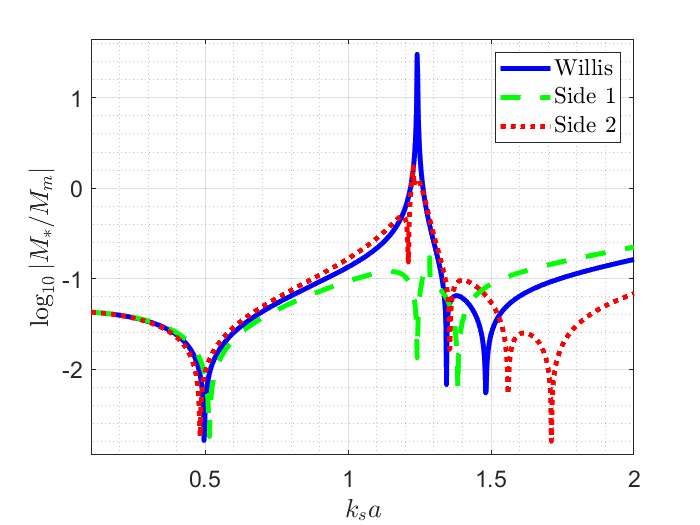}
\caption{}
\end{subfigure}
\hfill
\begin{subfigure}[h]{0.5\linewidth}
\includegraphics[width=\linewidth]{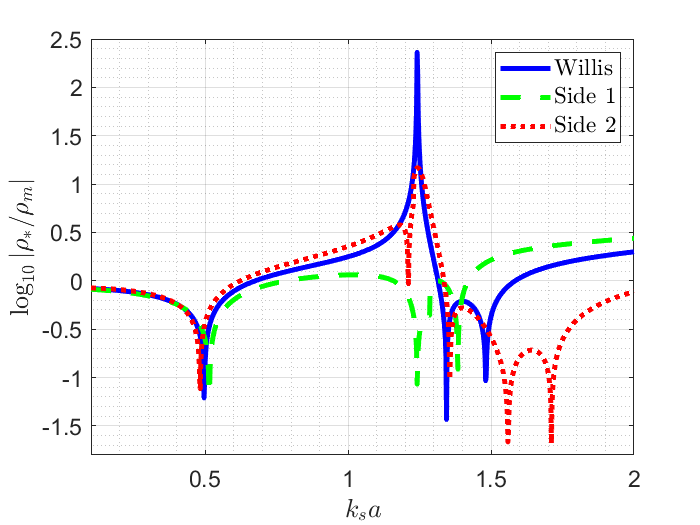}
\caption{}
\end{subfigure}\\
\begin{subfigure}[h]{1\linewidth}
\includegraphics[width=0.5\linewidth]{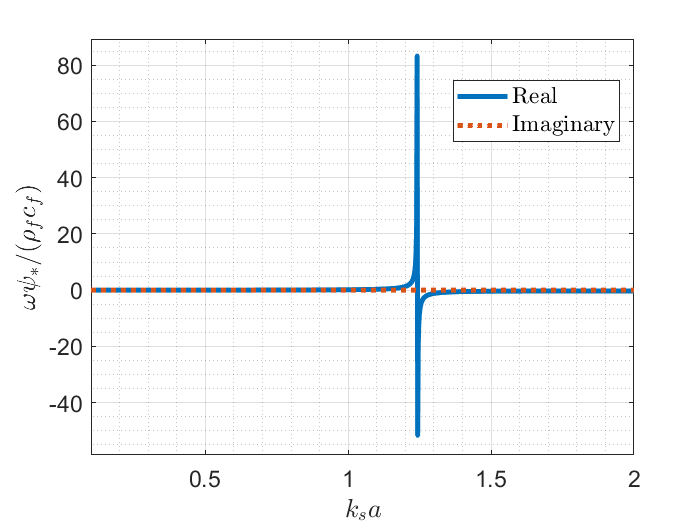}
\centering
\caption{}
\end{subfigure}
\caption{Plot of (a) $\log(M_*/M_m), (b) \log(\rho_*/\rho_m)$ and (c) $zW = \omega\psi_*/(\rho_fc_f)$ as functions of $k_s a (=k_ph)$. In (a) and (b): dashed - incidence from left without Willis coupling; dotted - incidence from right without Willis coupling; solid - incorporating both incidence fields, with Willis coupling.  Note, in (c), $zW$ is real as required for a passive material with no loss.
}
\label{nolossproperties}
\end{figure}

As a consequence of reciprocity, without loss we anticipate that $|R_1|=|R_2|$ with the only difference being in the phase of the reflection coefficients; this is illustrated in Fig.\ \ref{nolossRTphase} where we confirm this result. In Fig.\ \ref{nolossproperties} we plot the three effective properties associated with the meta-slab's reflection and transmission coefficients. We illustrate the effective compression modulus $M_*$ and density $\rho_*$ as deduced via single sided incidence, and also via double sided incidence incorporating Willis coupling. We then also show the effective Willis parameter, plotted as $zW$, that is required to ensure unique material properties. Note that effective properties are not uniquely defined if Willis coupling is not included.

\subsection{Meta-slab with loss}

In order to accommodate loss, which is a common effect in elastomeric media such as polymers, we introduce damping via the shear modulus, which becomes complex, through the damping parameter $\tan\delta=\Im(\mu_m)/\Re(\mu_m)$ and to ensure that the bulk modulus of the matrix medium remains real, we also introduce a small imaginary part to $M_m$. In the example here, we take $\tan\delta=0.1$ and for the compression modulus $\Im(M_m)/\Re(M_m) = 0.0013$. In reality the loss parameter will be frequency dependent \cite{neefjes2024stress} but to illustrate the effect and as is often done, we assume that it is constant here. 


\begin{figure}[h!]
\begin{subfigure}[h]{0.5\linewidth}
\includegraphics[width=\linewidth]{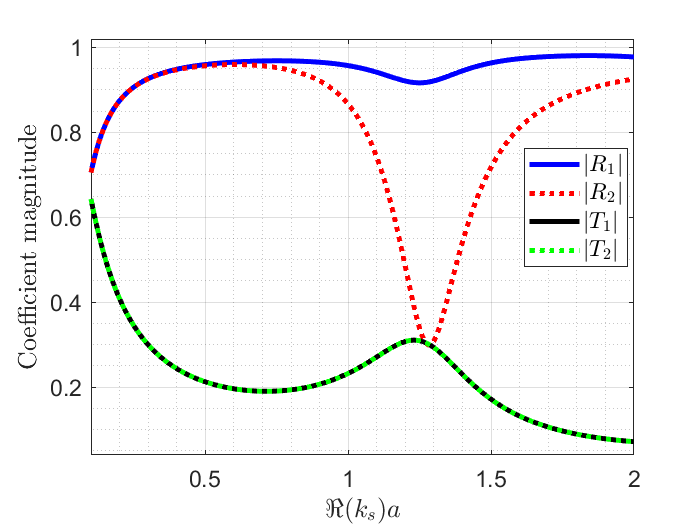}
\caption{}
\end{subfigure}
\hfill
\begin{subfigure}[h]{0.5\linewidth}
\includegraphics[width=\linewidth]{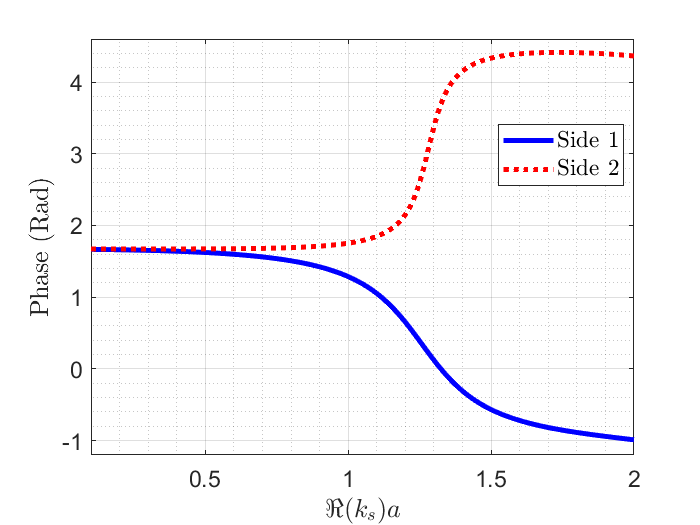}
\caption{}
\end{subfigure}
\caption{Plot of (a) $|R_1|, |R_2|, |T_1|$ and $|T_2|$ as functions of $\Re(k_s)a$, and (b) the phase of $R_1$ and $R_2$ relative to $T_1\backslash T_2$, again as functions of $\Re(k_s)a$, when there is loss in the matrix medium of the slab.
Illustrating the asymmetric reflection from the slab, indicative of the effect of Willis coupling. When the wave is incident on side 2 $(R_2)$ (small scatterers first) the amount of reflection is significantly lower than when the wave is incident on side 1 $(R_1)$ (large scatterers first) over a region close to $\mathcal{R}(k_s)a=1.3$
}
\label{includinglossRTphase}
\end{figure}

We retain the same configuration as above and in Fig.\ \ref{includinglossRTphase} illustrate the effect on reflection and transmission coefficients when waves are incident from either side of the meta-slab. Note the striking asymmetry effect here. Indeed, when incident from side 2 there is a huge reduction in the reflection coefficient just below $\Re(k_s)a=1.3$. The resonance is still present as can be seen from the equivalent curve for $|R_1|$, but the absorption is nowhere near as great. Incidence from side 2 is incidence on the side of the smaller voids. This appears to give rise to a longer propagation distance before significant scattering (from the larger voids) and thus a larger amount of dissipation close to the equivalent resonance frequency that arises due to the relatively close presence of the voids along the $x$ axis. 

\begin{figure}[h!]
\begin{subfigure}[h]{0.5\linewidth}
\includegraphics[width=\linewidth]{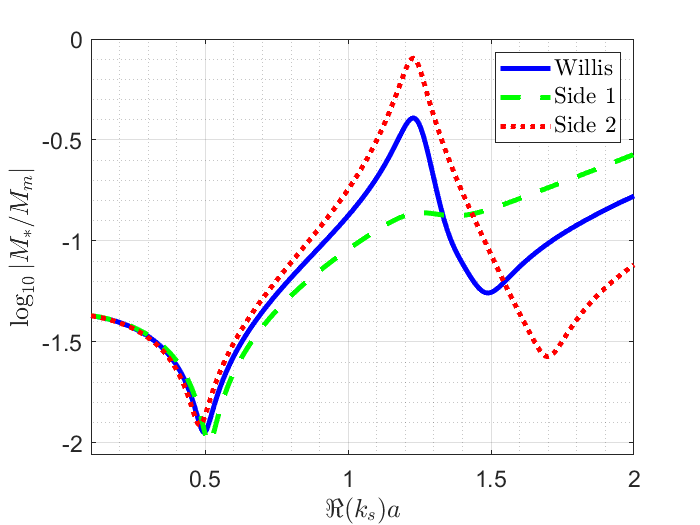}
\caption{}
\end{subfigure}
\hfill
\begin{subfigure}[h]{0.5\linewidth}
\includegraphics[width=\linewidth]{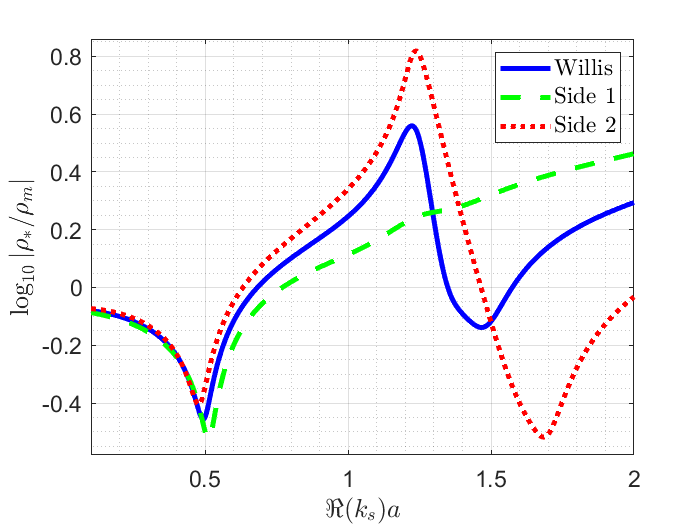}
\caption{}
\end{subfigure}\\
\begin{subfigure}[h]{1\linewidth}
\includegraphics[width=0.5\linewidth]{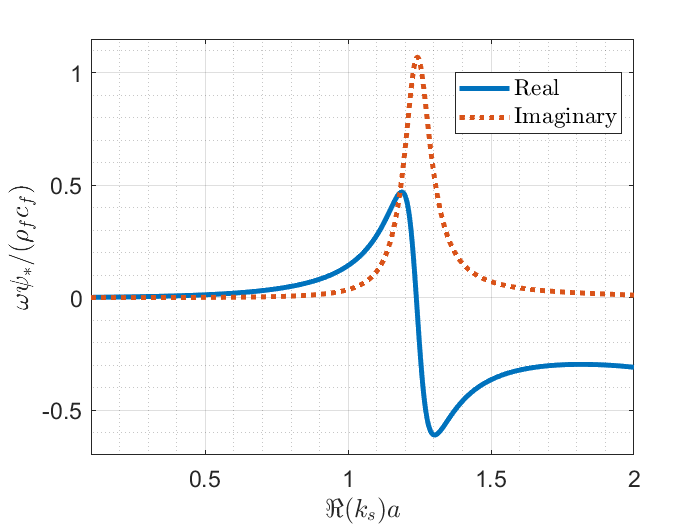}
\centering
\caption{}
\end{subfigure}
\caption{Plot of (a) $\log(M_*/M_m), (b) \log(\rho_*/\rho_m)$ and (c) $zW = \omega\psi_*/(\rho_fc_f)$ as functions of $\Re(k_s)a$. In (a) and (b): dashed - incidence from left without Willis coupling; dotted - incidence from right without Willis coupling; solid - incorporating both incidence fields, with Willis coupling.  Note, $zW$ is no longer real now that damping is incorporated within the slab.
}
\label{withlossproperties}
\end{figure}

It is interesting to note this asymmetry in absorption. We retain reciprocity here of course given that there is no energy source, but the local effective impedance conditions on the boundaries give rise to this asymmetric absorption effect. In Fig. \ref{withlossproperties} we illustrate the effective properties that are assigned to the meta-slab. Once again we note without the Willis coupling parameter non-unique parameters would be determined for the bulk modulus and density. When incorporating (complex) Willis coupling however we ensure uniqueness. 

\begin{figure}[h!]
\centering
\includegraphics[width=0.6\linewidth]{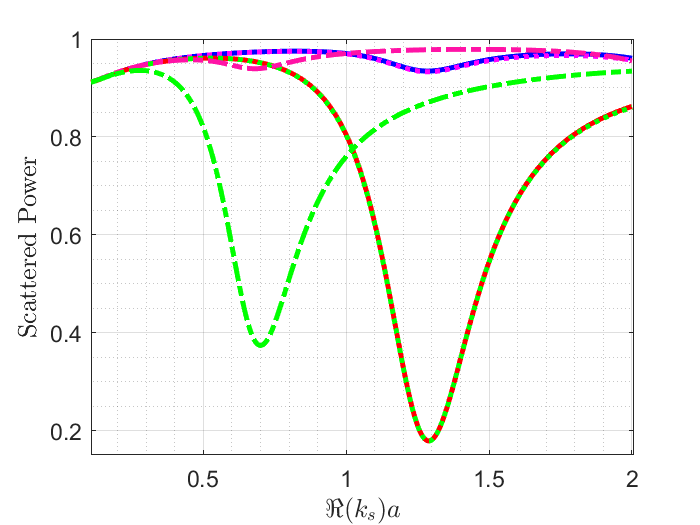}
\caption{Total scattered power comparing alternative configurations. $10a$ thickness slab with $3a$ void separation: solid blue (side 1) and solid Red (side 2); $20a$ thickness slab with $3a$ void separation: dotted pink (side 1) and dotted green (side 2). 
$20a$ thickness slab with $8a$ void separation: dot-dash pink (side 1) and dot-dash green (side 2). The larger separation in the thicker slab provides a lower frequency resonance whereas the $3a$ separation response is similar in both thick and thin slabs.}
\label{diffthicknessslabs}
\end{figure}

There are inevitably a vast range of different geometrical effects that one could study. Here however we illustrate some of these. Firstly, we consider the influence of thickening the slab, whilst retaining the $3a$ spacing between the voids. We do this by placing the void arrays at $x=9.4a$ and $x=12.4a$, still ensuring that the centre of mass of the slab is central. From Fig.\ \ref{diffthicknessslabs} we see that this however makes almost no difference to the meta-slab response. In this figure we plot the total scattered power, illustrating a key aspect of absorption of the slab. The results for $3a$ separation appear to be almost completely unchanged in the thick and thin meta-slabs. 

We consider an additional configuration in this figure in the thick ($20a$) meta-slab. We push the voids further apart, considering a spacing of $8a$ (with array centres located at $x=8.4a$ and $x=16.4$). From the figure we see that this appears to lower the effective resonance of the system, pulling the strong absorption on side 2 down to a lower frequency.  

\begin{figure}[h!]
\begin{subfigure}[h]{0.5\linewidth}
\includegraphics[width=\linewidth]{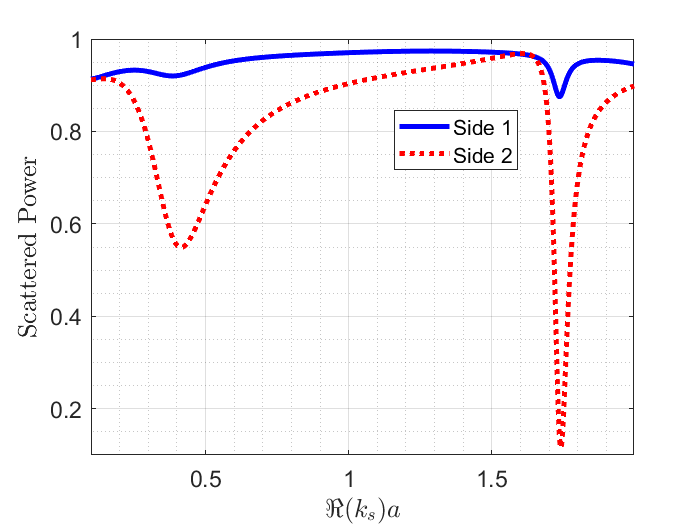}
\caption{}
\end{subfigure}
\hfill
\begin{subfigure}[h]{0.5\linewidth}
\includegraphics[width=\linewidth]{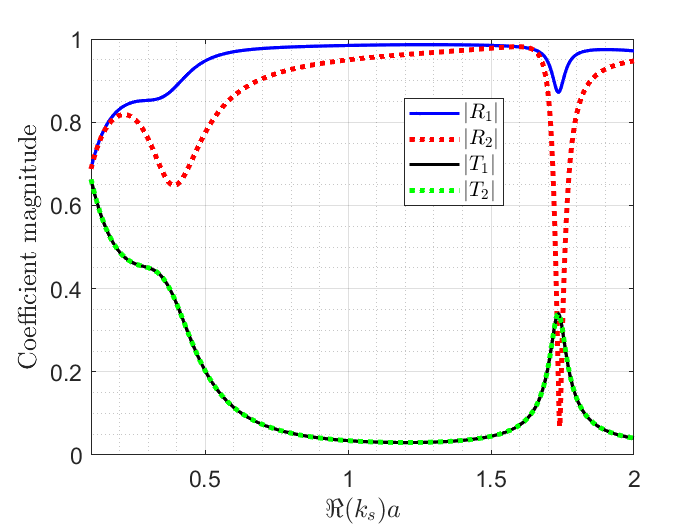}
\caption{}
\end{subfigure}
\caption{Response of a slab of thickness $80a$ and void separation $20a$, illustrating that in this instance the voids act as separate resonators with distinct resonances. (a) Scattered power and (b) respective reflection and transmission coefficients.
}
\label{thickcomps}
\end{figure}

Finally we consider an ultra-thick meta-slab, of thickness $80a$ and push the voids apart to a separation of $20a$. We see from Fig.\ \ref{thickcomps} that this leads to the voids now having distinct resonances. They are now acting in isolation to provide two different resonances to the reflection and transmission response of the meta-slab. 


\section{Conclusions} \label{Sec5}

We have shown that one can induce strongly directional-dependent absorption in an elastodynamic meta-slab and that such a medium can be represented in an effective manner via a Willis effective medium. Using classical effective elastic properties will not describe the medium adequately here due to the presence of asymmetric microstructure. The giant monopole resonance of the voids, due to the low rigidity parameter gives rise to this strong low frequency effect. Effective properties were deduced by coupling the reflection and transmission coefficients from an effective Willis medium to the specific scattering problem, with the latter being solved by a hybrid analytical-numerical scheme. 

A number of different configurations illustrated that one can tailor the resonant response significantly, with a strong dependence on the void radii, the spacing of the arrays and the slab thickness. 

This problem is the elastodynamic equivalent of the problem considered in Muhlestein et al. \cite{muhlestein2017experimental}. Significantly, although only the compressional modulus is deduced, shear plays a critical role in the response of the meta-slab via local multiple scattering in the slab, and the loss induced by his mechanism. Normal incidence means that we only couple into a single mode of the effective wave behaviour and thus we only deduce a single effective elastic constant and Willis coupling parameter. Oblique incidence allows further effective constants to be deduced however, which will be the subject of future work. Future work is also required to determine the link from microstructure to specific design/tailored effective properties and absorption characteristics.

\appendix

\section{Hybrid analytical-numerical multiple scattering method}

In the elastic slab we solve the full planar elastodynamic multiple scattering problem, and thus pose the classical series expansions of the field (via the two scalar potentials representing compression and shear).  These expansions represent both left and right going waves and must be matched to the acoustic fields at the slab's boundaries.  For the void boundary value problem, the elastic fields are written as sums of incoming and outgoing cylindrical modes, which are re-constituted into plane wave expansions at the slab boundaries.

In the following we consider an incident pressure wave propagating from the left but the adaptation to propagation from the right is obvious.  We begin by considering the impact of the periodic geometry of the void scattering system.

\subsection{Quasi-Periodic Systems}
The elastic slab configuration is illustrated in Fig \ref{config} for the symmetric case on the left, and the asymmetric (Willis) case on the right.  It lies in the region $0\le x \le h$, and the surface at $x=0$ and $x=h$ are in contact with acoustic half-spaces.  A plane pressure wave insonifies the slab's surface at $x=0$.  Within the slab are $N$ columns (here we restrict $N$ to be 1 or 2) of circular scatterers whose centres are located at $(\text{x}_i, \text{y}_i + jd)$, for $i = 1\to N$ and $j = -\infty \to \infty$, where $d$ is the separation between the centres of adjacent scatterers that lie within the same column.

The periodic spacing $d$, coupled with plane wave insonification, means that the resulting fields are quasi-periodic in $y$, that is, for any field variable $f$,
\begin{equation}
  f(x,y+jd) = e^{\text{i}j\tau}f(x,y),
\end{equation}
where $\tau$ represents the phase shift between adjacent cells of the periodic structure that is imposed by the form of the incident field.  For example if the slab is insonified by a plane acoustic wave propagating at an angle $\theta$ with respect to the slab normal then
\begin{equation}\label{eq:taudef}
  \tau = k_fd\sin\theta,
\end{equation}
where $k_f$ is the wavenumber of the incident pressure wave.

The fluid pressure, and elastic potential functions, are required to satisfy the Helmholtz equation.  Thus in $x < 0$, for example, the fluid pressure must be of the form
\begin{equation}\label{eq:fluid1}
  p_r(x,y) = p_i e^{\text{i}(\alpha_0 x + k_0 y)} + \sum_{j = -\infty}^\infty r_j e^{\text{i}(-\alpha_j x + k_j y)},
\end{equation}
where $r_j$ is the amplitude of the $j^{th}$ reflected, plane-wave mode, and $p_i$ is the amplitude of the incident field; further
\begin{equation}
  k_j = \frac{\tau + 2\pi j}{d}, \ \text{and } \alpha_j = \sqrt{k_f^2 - k_j^2},
\end{equation}
noting that $k_0 = \tau/d = k_f\sin\theta$ is the $y$-component of the wave-vector of the incident field; this form clearly satisfies the requirements of quasi-periodicity.
\subsection{Plane-wave expansion of the slab field}
When the incident pressure-wave hits the boundary of the elastic-slab at $x=0$, it generates a pair of compression and shear waves within the slab that propagate in the $+x$ direction with a common $y-$component, $k_0$, of their wave-vectors.  When these elastic-waves impinge upon the line of periodic scatterers they are scattered into all plane wave modes, $k_j$, in both the forward ($+x$) and backward ($-x$) directions.  These scattered waves propagate to the slab boundaries where they are specularly reflected (i.e. with the same $k_j$ component) into more compression and shear waves, which in their turn come back to interact with the scatterers.  At the slab boundaries, plane pressure-wave components are also transmitted into the fluid giving rise to the second term of \eqref{eq:fluid1}.  A similar form exists for the transmitted field in $x > h$ except the incident field is absent, that is
\begin{equation}\label{eq:fluid2}
  p_t(x,y) = \sum_{j = -\infty}^\infty t_j e^{\text{i}[\alpha_j (x-h) + k_j y]},\quad x > h.
\end{equation}

The elastic response of the slab is taken to be linear, with its displacements satisfying Navier's equations.  For this two dimensional problem, it is convenient to express the displacements in terms of the Helmholtz potentials $\varphi$ and $\psi$, viz
\begin{equation}
  \boldsymbol{u_s} = \frac{1}{k_s^2}\boldsymbol{\nabla}\times (\boldsymbol{\hat{z}}\psi) - \frac{1}{k_p^2}\boldsymbol{\nabla}\varphi,
\end{equation}
where $k_p$ and $k_s$ are respectively the compression and shear wavenumbers, and $\boldsymbol{\hat{z}}$ is a unit vector pointing in the $z-$direction (out of the $x-y$ plane of motion).  The two scalar potentials $\varphi$ and $\psi$ satisfy Helmholtz's equation
\begin{equation}
  (\nabla^2 + k_p^2)\varphi = 0,\quad\quad (\nabla^2 + k_s^2)\psi = 0.
\end{equation}

Imposing quasi-periodicity upon $\varphi$ and $\psi$ leads to the following representations of the elastic fields within the slab.  In solid region 1, defined as lying between the scatterers and the slab boundary at $x=0$,
\begin{equation}
  \varphi(x,y) = \sum_{j=-\infty}^\infty\left[A_j^+e^{\text{i}\gamma_j x} + A_j^-e^{-\text{i}\gamma_j x} \right] e^{\text{i}k_jy}, \quad x \in \text{region 1},
\end{equation}
\begin{equation}
  \psi(x,y) = \sum_{j-\infty}^\infty\left[C_j^+e^{\text{i}\beta_j x} + C_j^-e^{-\text{i}\beta_j x} \right] e^{\text{i}k_jy}, \quad x \in \text{region 1},
\end{equation}
whilst in solid region 2, defined as lying between the scatterers and the slab boundary at $x=h$,
\begin{equation}
  \varphi(x,y) = \sum_{j=-\infty}^\infty\left[E_j^+e^{\text{i}\gamma_j (x-h)} + E_j^-e^{-\text{i}\gamma_j (x-h)} \right] e^{\text{i}k_jy}, \quad x \in \text{region 2},
\end{equation}
\begin{equation}
  \psi(x,y) = \sum_{j-\infty}^\infty\left[F_j^+e^{\text{i}\beta_j (x-h)} + F_j^-e^{-\text{i}\beta_j (x-h)} \right] e^{\text{i}k_jy}, \quad x \in \text{region 2},
\end{equation}
where
\begin{equation}
  \gamma_j = \sqrt{k_p^2 - k_j^2},\quad \beta_j = \sqrt{k_s^2 - k_j^2}.
\end{equation}

In practice the infinite sums over plane wave modes, $j$, may be truncated at some suitable value, say $\pm J$, because once $|k_j|$ becomes greater than both $k_p$ and $k_s$ the compressional and shear waveforms are evanescent.  Thus higher order plane-wave modes decay rapidly with increasing $|x|$, and the higher order elastic wave modes that propagate away from the scatterers have negligible impact when they reach the slab boundaries.\footnote{Closer to the scatterers, the higher order modes may well be significant in field calculations but they can be disregarded in the initial boundary value problem.}

The connection between the incoming elastic plane waves, and the outgoing waves from the scatterers, may be calculated from multiple scattering theory.  For circular scatterers this can be achieved by decomposing each incident plane wave component, compression or shear, into a series of circular harmonics (an angular order decomposition).  The resulting scattered fields are obtained initially as angular order decompositions, which are then used to reconstruct the outgoing plane wave fields.  At slab boundaries, the boundary value is solved to obtain the relationships between the incident and reflected elastic waves, and the pressure waves transmitted into the fluid.\footnote{At the slab boundaries each incident Bloch/plane wave mode is specularly reflected and modes with different values of $k_j$ are decoupled.}  This gives rise to the following relationships.\newline
\newline
\textbf{Solid Region 1}

In solid region 1, the waves propagating in the $+x$ direction are a combination of the waves that would be generated by the incident pressure wave if the region $x>0$ were a homogeneous elastic half-space plus the reflections of the waves propagating in the $-x$ direction, that is
\begin{align}\label{eq:constr1}
  A_j^+ = R_j^{(11)}A_j^- + R_j^{(12)}C_j^- + a_0\delta_{0j},\\
  C_j^+ = R_j^{(21)}A_j^- + R_j^{(22)}C_j^- + c_0\delta_{0j},
\end{align}
where the matrices $\boldsymbol{R_j}$ are determined by the fluid-solid boundary value problem at $x=0$, and $a_0,c_0$ are the amplitudes of the `half-space' compression and shear waves.

The waves propagating in the $-x$ direction are the sum of the backscattered elastic waves of region 1 plus the forward scattered and incident waves of region 2, that is
\begin{align}
  A_j^- = &\sum_{l=-J}^{J}\left[Q_{jl}^{(11)}A_l^+ + Q_{jl}^{(12)}C_l^+ \right] + \nonumber\\
  &\sum_{l=-J}^{J}\left[ \left(\delta_{lj}e^{\text{i}\gamma_j h} + P_{jl}^{(11)}\right)E_l^- + P_{jl}^{(12)}F_l^-\right],
\end{align}
\begin{align}
  C_j^- = &\sum_{l=-J}^{J}\left[Q_{jl}^{(21)}A_l^+ + Q_{jl}^{(22)}C_l^+ \right] + \nonumber\\
  &\sum_{l=-J}^{J}\left[P_{jl}^{(21)}E_l^- + \left(\delta_{lj}e^{\text{i}\beta_j h}+ P_{jl}^{(22)}\right)F_l^-\right],
\end{align}
where the $\boldsymbol{Q}$ and $\boldsymbol{P}$ matrices are obtained from multiple scattering theory; they describe the scattering of plane-waves into region 1, ie waves scattered into the $-x$ direction by the circular scatterers. The $\boldsymbol{Q}$ matrix applies when the scatterers are insonified from region 1, whilst $\boldsymbol{P}$ matrix applies when the scatterers are insonified from region 2; the scattered plane-wave amplitudes are defined at $x=y=0$.
\newline
\newline
\textbf{Solid Region 2}

In solid region 2, similar considerations lead to
\begin{align}
  E_j^- = Z_j^{(11)}E_j^+ + Z_j^{(12)}F_j^+,\\
  F_j^- = Z_j^{(21)}E_j^+ + Z_j^{(22)}F_j^+,
\end{align}
where the matrices $\boldsymbol{Z_j}$ are determined by the fluid-solid boundary value problem at $x=h$.  And for the waves in region 2 that propagate in the $+x$ direction
\begin{align}
  E_j^+ = e^{\text{i}\gamma_j h}&\sum_{l=-J}^{J}\left[\left(\delta_{lj} + \tilde{Q}_{jl}^{(11)}\right) A_l^+ + \tilde{Q}_{jl}^{(12)}C_l^+ \right] + \nonumber\\
  e^{\text{i}\gamma_j h}&\sum_{l=-J}^{J}\left[\tilde{P}_{jl}^{(11)}E_l^- + \tilde{P}_{jl}^{(12)}F_l^-\right],
\end{align}
\begin{align}\label{eq:constr8}
  F_j^+ = e^{\text{i}\beta_j h}&\sum_{l=-J}^{J}\left[\tilde{Q}_{jl}^{(21)}A_l^+ + \left(\delta_{lj} + \tilde{Q}_{jl}^{(22)}\right)C_l^+ \right] + \nonumber\\
  e^{\text{i}\beta_j h}&\sum_{l=-J}^{J}\left[\tilde{P}_{jl}^{(21)}E_l^- + \tilde{P}_{jl}^{(22)}F_l^-\right],
\end{align}
where the $\boldsymbol{\tilde{Q}}$ and $\boldsymbol{\tilde{P}}$ matrices describe scattering of waves into the +$x$ direction.

Given the matrices $\boldsymbol{Q},\boldsymbol{\tilde{Q}},\boldsymbol{P}$ and $\boldsymbol{\tilde{P}}$, the system of equations defined by \eqref{eq:constr1} to \eqref{eq:constr8} can be solved for the scattering coefficients: $A_j^\pm, C_j^\pm, E_j^\pm$ and $F_j^\pm$.  The fluid reflection and transmission coefficients are then readily obtained from the fluid-structure boundary conditions on the slab sides.
\subsection{Multiple Scattering in an infinite, periodic system of circular scatterers}

For scattering from circular inclusions, the boundary value problem is best expressed in polar coordinates, where for the $j$'th scatterer, local polar coordinates $(r_j,\theta_j)$ are defined through $x = \text{x}_j + r_j\cos\theta_j, y = \text{y}_j + r_j\sin\theta_j$, recalling that $(\text{x}_j,\text{y}_j)$ denote the centre coordinates of the $j$'th scatterer within the global coordinate system.

The analysis for a finite number of circular scatterers, within an elastic medium, is described in \cite{cotterill2022deeply}.  Here, we consider $N$ columns of scatterers where within each column there are an infinite number of periodically spaced scatterers.  The scatterers in each column are identical, and have radius $a_n$ say.  The periodic spacing, $d$, is the same for all columns, thus the Cartesian coordinates at the centre of the $j$'th scatterer within the $n$'th column are $(\text{x}_n + jd,\text{y}_n + jd)$, where $(\text{x}_n,\text{y}_n)$ are the centre coordinates of the $n$'th base-cell void.

For this infinite system of scatterers, the total potential functions may be written \cite{martin2006multiple}
\begin{equation}\label{eq:comp1}
  \varphi(\boldsymbol{x}) = \varphi^{\text{in}}(\boldsymbol{x}) + \sum_{j=-\infty}^\infty\sum_{n=1}^N\sum_{m=-\infty}^\infty B_m^{(jn)}
  \text{H}_m^{(1)}(k_pr_{jn})e^{\text{i}m\theta_{jn}},
\end{equation}
\begin{equation}\label{eq:shear1}
  \psi(\boldsymbol{x}) = \psi^{\text{in}}(\boldsymbol{x}) + \sum_{j=-\infty}^\infty\sum_{n=1}^N\sum_{m=-\infty}^\infty D_m^{(jn)}
  \text{H}_m^{(1)}(k_sr_{jn})e^{\text{i}m\theta_{jn}},
\end{equation}
where $\varphi^{\text{in}}(\boldsymbol{x})$ and $\psi^{\text{in}}(\boldsymbol{x})$ denote the incident compression and shear fields respectively, and $B_m^{(jn)}$ and $D_m^{(jn)}$ are the corresponding scattering coefficients associated with the $j$'th scatterer of the $n$'th column at angular order $m$, with$(r_{jn},\theta_{jn})$ being the polar coordinates of the field point $(x,y)$ expressed within the local coordinate system of the $j$'th scatterer of the $n$'th column.

For plane-wave excitations the system is quasi-periodic which requires $B_m^{(jn)} = e^{\text{i}j\tau}B_m^{(0n)} = e^{\text{i}j\tau}B_m^{(n)}$, and $D_m^{(jn)} = e^{\text{i}j\tau}D_m^{(0n)} = e^{\text{i}j\tau}D_m^{(n)}$, allowing \eqref{eq:comp1} and \eqref{eq:shear1} to be written\footnote{For the slab problem considered here, $\tau$ was defined in \eqref{eq:taudef}, and has the same value for all compression and shear, plane-wave modes in the slab.}
\begin{equation}\label{eq:comp2}
  \varphi(\boldsymbol{x}) = \varphi^{\text{in}}(\boldsymbol{x}) +\sum_{n=1}^N\sum_{m=-\infty}^\infty B_m^{(n)}
  \sum_{j=-\infty}^\infty e^{\text{i}j\tau}\text{H}_m^{(1)}(k_pr_{jn})e^{\text{i}m\theta_{jn}},
\end{equation}
\begin{equation}\label{eq:shear2}
  \psi(\boldsymbol{x}) = \psi^{\text{in}}(\boldsymbol{x}) +\sum_{n=1}^N\sum_{m=-\infty}^\infty D_m^{(n)}
  \sum_{j=-\infty}^\infty e^{\text{i}j\tau}\text{H}_m^{(1)}(k_sr_{jn})e^{\text{i}m\theta_{jn}}.
\end{equation}

Close to the base cell scatterer $(j=0)$ of the $n$'th column, the total field may be formulated as
\begin{equation}
  \varphi(\boldsymbol{x}) = \sum_{m=-\infty}^\infty\left(\mathcal{A}_m^{(n)}\text{J}_m(k_pr_n) + B_m^{(n)}\text{H}_m^{(1)}(k_pr_n)\right)
  e^{\text{i}m\theta_n},
\end{equation}
\begin{equation}
  \psi(\boldsymbol{x}) = \sum_{m=-\infty}^\infty\left(\mathcal{C}_m^{(n)}\text{J}_m(k_sr_n) + D_m^{(n)}\text{H}_m^{(1)}(k_sr_n)\right)
  e^{\text{i}m\theta_n},
\end{equation}
where we have dropped the index $j=0$ for brevity of notation, and $\mathcal{A}_m^{(n)}$ and $\mathcal{C}_m^{(n)}$ are the model amplitudes of the \emph{total} incoming fields at the $n$'th scatterer, with the latter comprising the incident plane wave and the fields scattered by all the other voids.  $\mathcal{A}_m^{(n)}$ and $\mathcal{C}_m^{(n)}$ are obtained by expanding \eqref{eq:comp2} and \eqref{eq:shear2} within the local coordinate system of the $n$'th base cell scatterer, using Graf's addition theorem for all other voids.  For now, we consider the incident field to be a unit-amplitude, plane, compression wave, propagating at incidence angle $\theta_j = \arctan(k_j/\gamma_j)$, the angle of propagation of one of the compression plane-wave modes, that is
\begin{equation}\label{eq:pw_inc}
  \varphi^{\text{in}}(\boldsymbol{x}) = e^{\text{i}k_p(x\cos\theta_j + y\sin\theta_j)} = e^{\text{i}(\gamma_j x + k_j y)}.
\end{equation}
Expanding \eqref{eq:pw_inc} in the $n$'th base-cell coordinate system gives
\begin{equation}\label{eq:PW_inc2}
  \varphi^{\text{in}}(\boldsymbol{x}) = q_j^{(n)}\sum_{m=-\infty}^{\infty} \text{i}^m e^{-\text{i}m\theta_j}\text{J}_m(k_p r_n)e^{\text{i}m\theta_n},
\end{equation}
where $q_j^{(n)} = e^{\text{i}k_p(x_n\cos\theta_j + y_n\sin\theta_j)}$.  Inserting \eqref{eq:PW_inc2} into \eqref{eq:comp2} and \eqref{eq:shear2} (with $\psi^{\text{in}}(\boldsymbol{x}) = 0$), along with the corresponding expansion of the scattered fields (see e.g.\ \cite{linton2007resonant} for the case of a single, infinite column with scalar wave field) yields
\begin{equation}\label{eq:decom1}
  \mathcal{A}_m^{(n)} = q_j^{(n)}\text{i}^m e^{-\text{i}m\theta_j} + \sum_{l=-\infty}^\infty\left[B_l^{(n)}\mathfrak{S}_{l-m}(k_p) +
  \sum_{i\ne n}^{N}B_l^{(i)} S_{l-m}^{(ni)}(k_p)\right]
\end{equation}
\begin{equation}\label{eq:decom2}
  \mathcal{C}_m^{(n)} = \sum_{l=-\infty}^\infty\left[D_l^{(n)}\mathfrak{S}_{l-m}(k_s) + \sum_{i\ne n}^{N}D_l^{(i)}S_{l-m}^{(ni)}(k_s)\right],
\end{equation}
where
\begin{equation}
  \mathfrak{S}_l(\zeta) = \text{i}^l\sum_{u=1}^{\infty}\text{H}_l^{(1)}(u\zeta d)\left[(-1)^l e^{\text{i}u\tau} +  e^{-\text{i}u\tau}\right],\quad
  S_l^{(ni)}(\zeta) = \sum_{u=-\infty}^{\infty} e^{\text{i}u\tau} e^{\text{i}l\phi_n^{iu}} \text{H}_l^{(1)}(\zeta R_n^{iu}),
\end{equation}
%
%
with
\begin{equation}
  R_n^{iu} = \sqrt{(\text{x}_n - \text{x}_i)^2 + (\text{y}_n - \text{y}_i - ud)^2},\quad
  \phi_n^{iu} = \arctan\left(\frac{\text{y}_n - \text{y}_i - ud}{\text{x}_n - \text{x}_i} \right).
\end{equation}

The incoming and outgoing angular order coefficients for the $n$'th base-cell scatterer are related, at each angular order $m$ by the boundary problem for a single scatterer:
\begin{equation}\label{eq:Tmatrix}
  \begin{bmatrix}
    B_m^{(n)}\\
    D_m^{(n)}
  \end{bmatrix}
  = \boldsymbol{T_m^{(n)}}
  \begin{bmatrix}
    \mathcal{A}_m^{(n)}\\
    \mathcal{C}_m^{(n)}
  \end{bmatrix}
\end{equation}
which thus introduces the T-matrix for the system.

By combining the above with \eqref{eq:decom1} and \eqref{eq:decom2}, the global angular scattering coefficients, $B_m^{(n)}$ and $D_m^{(n)}$, can be found up to a suitable truncation point, say $\pm M$ in angular order, noting that $\boldsymbol{T_m^{(n)}}$ decays rapidly at large $m$ \cite{linton2007resonant}. The $\boldsymbol{T_m}$ matrix for voids was discussed in \cite{cotterill2022deeply} and the more general case of a rigid-in-soft (dipole) resonator was given in \cite{touboul2022enhanced}.

Given the coefficients $B_m^{(n)}$ and $D_m^{(n)}$, the fields scattered by the $N$ columns can be re-expressed as plane-wave expansions viz 
\begin{equation}
  \varphi^{(\text{sc})} = \sum_{l=-\infty}^{\infty} \mathfrak{A}_l^\pm e^{\text{i}(\pm\gamma_l x + k_l y)},\quad
  \psi^{(\text{sc})} = \sum_{l=-\infty}^{\infty} \mathfrak{C}_l^\pm e^{\text{i}(\pm\beta_l x + k_l y)},
\end{equation}
for waves propagating in the $\pm x-$directions, where
\begin{equation}
  \mathfrak{A}_l^\pm = \frac{2}{k_p d}\sum_{n=1}^{N}\sum_{m=-\infty}^{\infty} B_m^{(n)}\frac{e^{\mp\text{i}m\psi_l}}{\sin\psi_l}
  e^{-\text{i}(\pm\gamma_l x_n + k_l y_n)},
\end{equation}
\begin{equation}
  \mathfrak{C}_l^\pm = \frac{2}{k_s d}\sum_{n=1}^{N}\sum_{m=-\infty}^{\infty} D_m^{(n)}\frac{e^{\mp\text{i}m\alpha_l}}{\sin\alpha_l}
  e^{-\text{i}(\pm\beta_l x_n + k_l y_n)}.
\end{equation}
$\psi_l$ and $\alpha_l$ are the grazing angles of the scattered, compression and shear, plane-wave modes, measured clockwise from the $y-$axis, that is $\psi_l = \arctan(\gamma_l/k_l)$ and $\alpha_l = \arctan(\beta_l/k_l)$.  Recalling that the above analysis was for a unit-amplitude plane compression wave propagating at an incidence angle of $\theta_j$ from the slab boundary at $x=0$, the definitions of the $\boldsymbol{Q}$ and $\boldsymbol{\tilde{Q}}$ matrices show that
\begin{equation}
  Q_{lj}^{(11)} = \mathfrak{A}_l^-,\quad Q_{lj}^{(21)} = \mathfrak{C}_l^-,\quad \tilde{Q}_{lj}^{(11)} = \mathfrak{A}_l^+,\quad \tilde{Q}_{lj}^{(21)} = \mathfrak{C}_l^+.
\end{equation}
To obtain the other terms of the $\boldsymbol{Q}$ and $\boldsymbol{\tilde{Q}}$ matrices, we remove the incident compression wave in \eqref{eq:comp2} and replace it with an incident shear wave in \eqref{eq:shear2}.  And to obtain the coefficients of the $\boldsymbol{P}$ and $\boldsymbol{\tilde{P}}$ matrices, we repeat the process with compression and shear waves propagating in the negative $x-$direction from the slab boundary at $x=h$, which, for the compression wave example above, is equivalent to replacing $\theta_j$ with $\pi - \theta_j$, and setting its amplitude to unity on $x=h$.

\enlargethispage{20pt}\vskip6pt


\vskip2pc

\bibliography{manuscript}
\bibliographystyle{plain}



\end{document}